\documentclass[12pt,preprint]{aastex}
\usepackage{graphicx,epsfig,epsf,amssymb}

\begin{document}

\newcommand{\logh}{$+5\log h$}
\newcommand{\mipsmu}{$24\mu$m}
\newcommand{\mpc}{\,{\rm Mpc}}
\newcommand{\kpc}{\,{\rm kpc}}
\newcommand{\pc}{\,{\rm pc}}
\newcommand{\magsec}{\,{\rm mag\,arcsec^{-2}}}
\newcommand{\Dmin}{{\,D_{\rm min}}}
\newcommand{\rmin}{{\,R_{\rm min}}}
\newcommand{\hkpc}{$h^{-1}$~kpc}
\newcommand{\beqn}{\begin{equation}}
\newcommand{\eeqn}{\end{equation}}
\newcommand{\sigone}{$\Sigma_{1}$}

\def\liu#1{{\bf #1}}

\shortauthors{Liu et al.}
\shorttitle{The UV-optical color gradients in SFGs at $\lowercase{z}\sim1$}
\title{The UV-optical Color Gradients in Star-Forming Galaxies at $0.5<\lowercase{z}<1.5$: 
Origins and Link to Galaxy Assembly}
\author{F. S. Liu \altaffilmark{1,2}, 
Dongfei Jiang \altaffilmark{1,3}, 
Yicheng Guo \altaffilmark{2}, 
David C. Koo \altaffilmark{2}, 
S. M. Faber \altaffilmark{2}, 
Xianzhong Zheng \altaffilmark{3}, 
Hassen M. Yesuf \altaffilmark{2}, 
Guillermo Barro \altaffilmark{2,4}, 
Yao Li \altaffilmark{1}, 
Dingpeng Li \altaffilmark{1}, 
Weichen Wang \altaffilmark{5}, 
Shude Mao\altaffilmark{5}, 
and Jerome J. Fang\altaffilmark{2,6}
}

\email{Email: fsliu@synu.edu.cn}

\affil{
$^1$College of Physical Science and Technology, Shenyang Normal University, Shenyang 110034, China\\
$^2$
University of California Observatories and the Department of Astronomy and Astrophysics, University of California, Santa Cruz, CA 95064, USA \\
$^3$Purple Mountain Observatory, Chinese Academy of Sciences, 2 West-Beijing Road, Nanjing 210008, China\\
$^4$Department of Astronomy, University of California, Berkeley, CA 94720-3411, USA\\
$^5$Department of Physics, Tsinghua University, Beijing 100084, China\\
$^6$Orange Coast College, Costa Mesa, CA 92626, USA
}

\begin{abstract}

The rest-frame UV-optical (i.e., $NUV-B$) color index is sensitive to 
the low-level recent star formation and dust extinction, but it is insensitive to the metallicity. 
In this Letter, we have measured the rest-frame $NUV-B$ color gradients 
in $\sim1400$ large ($\rm r_e>0.18^{\prime\prime}$), nearly face-on ($b/a>0.5$) main-sequence star-forming galaxies (SFGs) 
between redshift 0.5 and 1.5 in the CANDELS/GOODS-S and UDS fields. With this sample, we study the origin of UV-optical 
color gradients in the SFGs at $z\sim1$ and discuss their link with the buildup of stellar mass. 
We find that the more massive, centrally compact, and more dust extinguished SFGs 
tend to have statistically more negative raw color gradients (redder centers) 
than the less massive, centrally diffuse, and less dusty SFGs. 
After correcting for dust reddening 
based on optical-SED fitting, the color gradients in the low-mass ($M_{\ast} <10^{10}M_{\odot}$) SFGs 
generally become quite flat, while most of the high-mass ($M_{\ast} > 10^{10.5}M_{\odot}$) SFGs 
still retain shallow negative color gradients. These findings imply that dust reddening is likely 
the principal cause of negative color gradients in the low-mass SFGs, 
while both increased central dust reddening and buildup of compact old bulges 
are likely the origins of negative color gradients in the high-mass SFGs. 
These findings also imply that at these redshifts the 
low-mass SFGs buildup their stellar masses in a self-similar way, 
while the high-mass SFGs grow inside out.

\end{abstract}

\section{Introduction} 

Investigating the spatial distributions of galaxy properties (i.e., color, stellar mass, 
star formation rate, etc.) is a powerful way to understand the buildup 
and shutdown of galaxies. Color gradient 
can provide information on the stellar population within a galaxy, through 
its dependency on the galactic distributions of dust, stellar age and metallicity \citep[][]{Gonzalez-Perez+11,Guo11}. 
In the local universe, early-type galaxies (ETGs) 
are generally old and dead, containing relatively little or no dust and usually have 
negative optical and near-infrared color gradients (redder centers). 
Previous studies \citep[e.g.,][]{Wu+05,Tortora+10} have shown that 
metallicity and not age gradients are the principal origin of 
such color gradients. 
Although color gradients in nearby late-type galaxies (LTGs) appear to be dominated by the fact that, 
on average, their bulges are redder than their disks, LTGs  show age, metallicity and dust gradients, due to the interplay among dust content, 
stellar feedback, bursts of star formation and stellar migration \citep[][]{Gonzalez-Perez+11}. 
Earlier studies have claimed a weak correlation of color gradients 
with the physical properties of local galaxies
\citep[e.g. mass, luminosity, etc.,][]{Peletier90,Kobayashi99,Tamura+03},
Recent studies on the basis of larger and homogeneous samples of galaxies in SDSS have found that color gradients 
in local galaxies are closely correlated with their stellar masses, luminosities, sizes, 
and residual star formation \citep[e.g.,][]{Tortora+10,Suh+10,Pan+15}.
These results have provided important information on the physical origin of
color gradients in local galaxies and clues on their buildup and shutdown. 

At moderate and high redshifts, the relationship between color gradient and galaxy properties 
have not been fully explored to date. By stacking the {\it HST} multi-band imaging, 
Wuyts et al. (2012) studied the resolved colors and stellar populations 
of a few hundred star-forming galaxies (SFGs) with $M_{\ast} >10^{10}$$M_{\odot}$ at $0.5<z<2.5$. 
They found evidence for redder color, lower star formation rate (SSFR), and 
increased dust extinction in the centers of galaxies, which are consistent with an 
inside-out disk growth scenario in massive SFGs. 
In a series of papers by Nelson et al. (i.e., Nelson et al. 2012,2015,2016), they studied the spatial map of SSFR traced by $H{\alpha}$ 
and dust map in SFGs at moderate redshifts ($z\sim1$ and $z\sim1.4$). 
Nelson et al. (2012) showed that the $H{\alpha}$ sizes of massive galaxies 
are bigger than their sizes in the rest-frame R-band. 
Nelson et al. (2015) showed that the $EW(H{\alpha})$ is flat with radius for 
the low-mass ($\rm log M_{\ast}/M_{\odot}=9.0-9.5$) galaxies, while the $EW(H{\alpha})$ decreases by a factor
of $\sim2$ on average from center to $2r_e$ for high mass galaxies.
These findings suggested that massive SFGs build up their stellar masses from inside out, 
while the low-mass SFGs grow in a self-similar way, irrespective of the radial distance. 
Note that dust correction was not done in these two works. So the results may 
be  due to either lower central SSFR or higher central dust or combination of both in massive galaxies.
Nelson et al. (2016) corrected for dust-extinction by 
using the Balmer decrement ($H{\alpha}/H{\beta}$). 
As a result, central dust was found as a major factor in causing radial color gradient 
in galaxies with a mean mass of $\rm \langle log M_{\ast}/M_{\odot} \rangle =10.2$. Galaxies 
with $\rm \langle log M_{\ast}/M_{\odot} \rangle =9.2$ have little dust attenuation at any radii. 
These studies suggest that both dust and stellar population changes can result 
in a significant radial color gradient in a SFG at moderate redshift. 

To a large extent, the degeneracies among the effects of dust, stellar age and metallicity 
on color gradients are hard to disentangle without independent information. 
However, metallicity usually affects the color of redder stars more strongly than 
the color of bluer ones. 
So it has been demonstrated successfully that, for the study of color gradients in 
local ETGs with little dust, including rest-frame near-infrared data is helpful to 
break the age-metallicity degeneracy (e.g., Wu et al. 2005). 
On the other hand, it has also been shown that the 
rest-frame UV-optical color index is sensitive to the low-level recent star formation (i.e., SSFR) 
and dust extinction, but is insensitive to the metallicity (e.g., Kaviraj et al. 2007; Pan et al. 2015; 
Fang et al. in prep.). In this Letter, we exploit deep high-resolution {\it HST}/WFC3 
and ACS multi-band imaging data to measure 
the rest-frame $NUV-B$ color gradients for a sample of $\sim1400$ SFGs 
near the ridge-line of the main sequence (MS), 
with $\rm log M_{\ast}/M_{\odot} >10^{9}$ at $0.5 < z < 1.5$ selected from the CANDELS/GOODS-S and UDS fields. 
These data allow us to make a statistically robust analysis of UV-optical color gradients in actively SFGs 
at moderate redshifts. We examine the effects of dust and bulge formation on color gradients in SFGs 
and further discuss the link between color gradient and the stellar mass assembly.

Throughout the Letter, we adopt a cosmology with a matter density 
parameter $\Omega_{\rm m}=0.3$, a cosmological constant 
$\Omega_{\rm \Lambda}=0.7$ and a Hubble constant of ${\rm H}_{\rm 
0}=70\,{\rm km \, s^{-1} Mpc^{-1}}$. All magnitudes are in the AB 
system. 

\section{DATA}

The sample of galaxies used in this work is selected from the first two 
available fields (GOODS-S \& UDS) of the CANDELS \citep[][]{Grogin+11,Koekemoer+11}. 
Based on the source detection in the WFC3/F160W band, the CANDELS team 
has made a multi-wavelength catalog for each field, 
combining the newly obtained CANDELS {\it HST}/WFC3 data with existing public ground-based and 
space-based data. {\it HST} photometry was measured by running {\tt SExtractor} \citep[][]{Bertin96} 
on the point spread function (PSF)-matched images in the dual-image mode, 
with the F160W image as the detection image. Photometry in ground-based and IRAC images, 
whose resolutions are much lower than that of the F160W images, was 
measured by using {\tt TFIT} \citep[][]{Laidler07}, 
which fit the PSF-smoothed high-resolution image templates to the low-resolution images 
to measure the fluxes in the low-resolution images. We refer readers to 
\citet[][]{Guo+13} and \citet[][]{Galametz13} for details on these data and reduction.

Photometric redshifts were estimated from a variety of different codes available 
in the literature, 
which are then combined to improve the individual performance \citep[][]{Dahlen13}. 
Rest-frame total magnitudes were computed from the
best available redshifts (spectroscopic or photometric) and multi-wavelength
photometry using {\tt EAZY} \citep[][]{Brammer+08}.
Stellar masses were computed using {\tt FAST} \citep[][]{Kriek+09} and based on a grid of \citet[][BC03]{BC03} models
that assume a \citet[][]{Chabrier03} IMF, solar metallicity, exponentially declining
star formation histories, and a \citet[][]{Calzetti00} dust extinction law. %
SFRs were computed by combining IR and rest-frame UV (uncorrected for dust extinction) 
luminosities (Kennicutt 1998 and Bell et al. 2005) and adopting a Chabrier 
IMF: $\rm SFR_{IR+UV}=1.09\times10^{-10}(L_{IR}+3.3L_{2800})$. 
Total IR luminosities ($\rm L_{IR}{\equiv}L[8-1000{\mu}m]$) were derived from Chary \& Elbaz (2001) 
templates fitting {\it MIPS} $24{\mu}m$ fluxes. 
For galaxies undetected by {\it MIPS} below a $\rm 2\sigma$ level ($\rm 20{\mu}Jy$), 
SFRs come from rest-frame UV luminosities at $\lambda\approx2800\AA$ 
that are corrected for extinction by assuming a Calzetti law ($\rm A_{2800}\approx1.79A_V$) 
with the median $A_V$ from all methods in \citet[][]{Santini+15}. 
Effective radius along the major axis ($\rm r_e$) 
was measured from the {\it HST}/WFC3 F160W and F125W 
images respectively using {\tt GALFIT} \citep[][]{Peng+02} 
and PSFs created and processed to replicate the conditions of the observed data \citep[][]{vdWel+12}.

The {\it HST} based multi-wavelength and multi-aperture photometry catalogs 
with improved local background subtraction were built for galaxies in the CANDELS fields 
(Liu et al. in prep.), which include the radial profiles of 
observed surface brightness and cumulative magnitude 
in the {\it HST}/WFC3 (F105W, F125W F140W, and F160W) bands and {\it HST}/ACS 
(F435W, F606W, F775W, F814W and F850LP) bands if available. 
Part of these data has been used in recent works by Barro et al. (2015a,b). 
For this work, we computed the rest-frame $NUV$ and $B$ band surface brightness profiles, 
stellar mass profiles, and the profiles of other stellar population 
parameters (i.e., $\rm A_V$, UV-based SFR, etc.) by fitting the best-fit SEDs in each photometry bin. 
The modeling is also based on a grid of BC03 models that assume a Chabrier IMF, 
solar metallicity, exponentially declining star formation histories, 
and a Calzetti extinction law. 
Note that the reddest filter in our radial data is limited to H(F160w). 
An extensive study by Wang et al. (in prep.) will establish that 
reasonable $\rm A_V$ values can be derived out to $z\sim1.5$. 
The multi-band {\it HST} mosaics were PSF-matched to the resolution of
F160W that has a half width at half maximum of HWHM=0.09$^{\prime\prime}$,
which corresponds to $\sim0.73$~kpc on average in our redshift range.
In order to compensate for the light smeared outside 1~kpc of galaxies due to PSF smoothing,
we followed Barro et al. (2015b) to 
add a S\'ersic dependent correction to derived central stellar mass surface 
density within a radius of 1~kpc ($\rm \Sigma_{1}$).

\section{Sample Selection}

The full GOODS-S and UDS catalogs contain 34,930 and 35,932 objects, respectively. 
The sample used in this work is constructed by applying the following cuts to the above data:

1. Observed F160W magnitude $\rm H < 24.5$ and the quality $\rm flag = 0$ 
(van der Wel et al. 2012) to ensure well-constrained {\tt GALFIT} measurements 
and eliminate doubles, mergers, and disturbed objects.

2. Photometry quality flag $\rm PhotFlag = 0$ to exclude spurious sources

3. SExtractor $\rm CLASS\_STAR < 0.9$ to reduce contamination by stars

4. Redshifts within $0.5<z<1.5$ and stellar masses at $\rm logM_{\ast}/M_{\odot}>9.0$ 
to maintain mass completeness 

5. Well-constrained measurements of surface brightness profiles 
at least in two of {\it HST}/ACS bands and two of {\it HST}/WFC3 bands 
simultaneously to guarantee the accuracy of best-fit SEDs

6. Axis ratio $b/a>0.5$ to remove galaxies with significant 
interplay of dust reddening and stellar population changes

7. $\rm {r_e}>0.18^{\prime\prime}~(3~pixels)$ to minimize the effect of 
PSF-matching on color gradient measurement

After the above cuts, we obtain 1,905 galaxies: 1,008 from GOODS-S and 897 from UDS. 
We then utilize rest-frame $UVJ$ diagram ($(U-V) > 0.88 \times (V-J)+0.49$, Williams et al. 2009) 
to select 1,606 SFGs (see left panels in Figure 1). Furthermore, we follow the method of 
Fang et al. (in prep.) to select galaxies near the ridge-line of 
the star-forming main-sequence (SFMS). Right panels in Figure 1 show the SSFR-mass relation 
for selected galaxies. Only galaxies defined as star-forming by the 
$UVJ$ criterion are included. As can be seen, the distributions of points
clearly trace out the SFMS at these redshifts (e.g., Whitaker et al. 2012 and reference therein), 
while the ``green-valley'' galaxies appear as the tails of objects below the MS. 
The solid lines indicate the adopted best-fit linear relation ($\rm logSSFR=-0.093(logM_{\ast}-10)-8.857)$ 
to galaxies in our redshift range, after excluding ``green-valley'' galaxies, defined 
as galaxies with vertical offsets from the best-fit relation of 
$\rm \triangle log SSFR < -0.45~dex$ (below dashed lines). 
Our choice to exclude such ``green-valley'' objects is due to 
significant decrease in their number at low mass galaxies after 
the size and axis-ratio cuts, which is not sufficient for a statistically
robust analysis. 
We will study these galaxies in the future with a large sample 
assembled from all five CANDELS fields.
The present paper focuses on the 1,432 the main-sequence SFGs. 

\section{Measurements of Color Gradients}

Radial surface brightness profiles along the major axis in all available {\it HST} bands 
for our galaxies were obtained. Rest-frame $NUV$ and $B$ band surface brightness profiles 
were computed using {\tt EAZY} \citep[][]{Brammer+08}. A linear least-squares method 
was then adopted to fit the derived $NUV-B$ color profiles. To reduce PSF effect, 
an inner radius of 0.''09 cut was used when fitting. As the data became quite noisy at large radius, 
the outer radius was cut at $2r_e$ to ensure errors lower than 0.1 mag $arcsec^{-2}$ 
in all bands. Consistent with previous studies (e.g., Peletier et al. 1990; Wu et al. 2005), 
the color gradient was expressed as $\nabla{(NUV-B)} \equiv d(NUV-B)/d log(r)$. 
The color gradients computed by this definition are tightly correlated with 
color difference between inner ($r<r_e$) and outer ($r_e<r<2r_e$) parts. We prefer such 
definition because it can reflect intrinsic slope in a color profile when considering the difference 
in radius. Figure 2 illustrates our measurement for an example galaxy (GOODS-S 16617). 
The color gradient in this example galaxy has a negative value (-0.37), which indicates 
a redder center (vice versa). The typical uncertainty 
in our color gradients is $\rm \sim0.1~dex$, which includes 
the fitting error and the errors of photometry (i.e., readout noise, sky subtraction, 
and PSF matching, etc.). 

\section{Results and Analysis}

Figure 1 presents the rest-frame $UVJ$ diagram (left panels) and SSFR-mass relation (right panels) 
for our MS SFGs at $0.5<z<1.5$, which 
are shown with points color-coded by the global $\rm A_V$ (top panels) 
and $\rm {log \Sigma_{1}}$ (bottom panels), respectively. 
In this plot, we can find the following information:

1. Bottom-right panel shows that $\rm{\Sigma_{1}}$ and stellar mass 
have a tight positive correlation (also see Barro et al. 2015b), which implies that 
more massive galaxies may harbour larger bulges and thus tend to 
have denser centers, since $\rm{\Sigma_{1}}$ is a powerful parameter to 
quantify the bulge growth as galaxies evolve \citep[e.g.,][]{Fang+13}. 

2. Fang et al. (in prep.) have found that the reddening contours 
in $UVJ$ diagram run nearly vertically for MS SFGs, as shown in our top-left panel, which demonstrates that
reddening correlates most strongly with $V-J$. The bottom-left panel indicates that 
more centrally compact galaxies are globally more dust extinguished. 

3. More centrally compact galaxies are globally redder in both $U-V$ and $V-J$, 
likely due to a combination of increased dust reddening 
and buildup of larger, older bulges in the centers. 

The upper panels in Figure 3 replot the $UVJ$ diagram (left) and 
SSFR-mass relation (right) for our galaxies, but this time the data 
are binned and color-coded by raw dust-reddened $NUV-B$ color gradients 
as observed ($\nabla_{obs}{(NUV-B)}$). Direct correlations with 
stellar masses, $\rm{\Sigma_{1}}$, and rest-frame $V-J$ colors are shown in 
the upper panels in Figure 4 (color-coded by stellar mass). 
We find that the MS SFGs at $0.5<z<1.5$ generally have negative $NUV-B$ color gradients (redder centers). 
The correlations with stellar mass and $\rm{\Sigma_{1}}$ 
imply that more massive and centrally compact galaxies
tend to have more negative gradients. The correlation 
with $V-J$ indicates that galaxies also tend to 
have more negative color gradients with increased dust reddening. 

In order to further understand the origin of negative $NUV-B$ 
color gradients observed in these actively SFGs, we attempt to remove dust effect by 
making a correction for derived $NUV-B$ color profiles and re-compute 
the dust-corrected gradients ($\nabla_{dc}{(NUV-B)}$).  
This correction exploits derived $\rm A_V$ profiles by fitting the optical-SEDs 
in each photometry bin and also assumes a Calzetti law ($\rm A_{NUV}\approx1.79A_V$ 
and $\rm A_{B}\approx1.26A_V$). 
The results are shown in the bottom panels in Figure 3 and Figure 4. 
It can be seen, after our correction for dust reddening, 
rest-frame $NUV-B$ color gradients in the low-mass ($M_{\ast} <10^{10}M_{\odot}$) galaxies 
generally become quite flat, which implies that dust reddening is likely the principal cause 
of negative $NUV-B$ color gradients in the low-mass galaxies. 
The neutral dust-corrected $NUV-B$ color gradients also imply that these low-mass galaxies 
are building up their stellar masses in a self-similar way, irrespective of the radial distance. 
However, most of the high-mass ($M_{\ast} > 10^{10.5}M_{\odot}$) galaxies still 
retain shallow negative $NUV-B$ color gradients after correcting for dust. The buildup of 
compact, old bulges in their centers is likely responsible for 
such residual color gradients. Therefore, both increased dust reddening and 
presence of compact, old bulges in the centers are likely the origins of negative 
$NUV-B$ gradients in the high-mass galaxies. The negative 
dust-corrected $NUV-B$ color gradients in these high-mass galaxies 
are consistent with an inside-out growth scenario. 

\section{Discussion and Conclusions}

In this Letter, we have measured the rest-frame $NUV-B$ color gradients in 
$\sim1400$ large ($\rm r_e>0.''18$), face-on ($b/a>0.5$) SFGs 
near the ridge-line of the MS between redshift 0.5 and 1.5 in 
the CANDELS/GOODS-S and UDS fields. With this sample, 
we for the first time make a statistically robust analysis of UV-optical color gradients 
in actively SFGs at $z\sim1$. Since the rest-frame UV-optical color index is sensitive to 
SSFR and dust extinction, but insensitive to the metallicity, we corrected for dust extinction 
to disentangle the effects. Our main conclusions are as follows:

1. Both massive and centrally compact SFGs are globally more dust extinguished.

2. More massive, centrally compact and more dust extinguished SFGs 
tend to have more negative raw $NUV-B$ color gradients than less massive, 
centrally diffuse and less dusty SFGs.

3. After correcting for dust reddening, the $NUV-B$ 
color gradients in the low-mass ($M_{\ast} <10^{10}M_{\odot}$) SFGs 
generally become quite flat, while most of the high-mass ($M_{\ast} > 10^{10.5}M_{\odot}$) SFGs 
still retain shallow $NUV-B$ negative gradients. These findings imply that dust reddening is likely
the principal cause of negative color gradients in the low-mass SFGs, whereas 
both increased dust reddening and buildup of compact, old bulges in the centers 
are likely the origins of negative color gradients in the high-mass SFGs. 

4. The neutral dust-corrected $NUV-B$ color gradients in the low-mass SFGs 
imply that their stellar masses grow self-similarly at all radii. 
The negative dust-corrected $NUV-B$ color gradients in the high-mass SFGs 
imply that central stars form earlier than outer stars, i.e., that 
massive galaxies are growing from inside out. 

Recently, van der Wel et al. (2014) showed that the low-mass galaxies at these redshifts 
are not disk-like but are prolate and irregular, unlike massive galaxies, 
which are shaped like normal, oblate disks. A consistent picture with our data 
is emerging in which young, low-mass galaxies first grow in an irregular manner, 
where new generations of stars are randomly mixed with previously existing populations, and that 
inside-out growth only occurs once galaxies attain a sustained, disk-like structure. 
Assuming that our fourth conclusion persists over a substantial range 
of redshifts, we infer that star-forming galaxies start out their lives 
growing self-similarly in their early phases and switch to growing from 
inside out when their masses pass $\rm M_{\ast} \sim 10^{10} M_{\odot}$. 
Since centers appear redder (older) 
after that, the change must happen because the centers slow down rather than 
the reverse (having the outer regions accelerate). This change marks 
the entry into the later stages of star-formation and signals incident 
entry into the transition region (green valley), followed by quenching. 
It will be useful to use galaxy formation models to put actual time markers 
on these various stages.

We stress that major conclusions in this paper depend on the SED modeling 
assumptions applied to the CANDELS data, namely (1) that galaxy stellar populations are single $\tau$-models, 
(2) that the dust extinction law is assumed to be the Calzetti, and 
(3) that the solar metallicity applies to all galaxies. 
These standard assumptions are what almost all of high-z studies are currently using. 
This paper does not attempt to justify these current state of the art assumptions, but
take the standard assumptions as given and aims to see where they lead.
%
%
We refer to the reader to Fang et al. (in prep.) about the 
uncertainties introduced by these standard assumptions. 
Future works should investigate the consequences 
of more realistic stellar population models, metallicity, and extinction law.

\acknowledgments

We thank Chenggang Shu and Zhu Chen for useful discussions. 
We acknowledge the anonymous referee for a constructive report that significantly 
improved this paper.
This project was supported by the NSF grants of China (11103013, 11573017).

\begin{figure*}
\centering
\includegraphics[angle=0,width=0.95\textwidth]{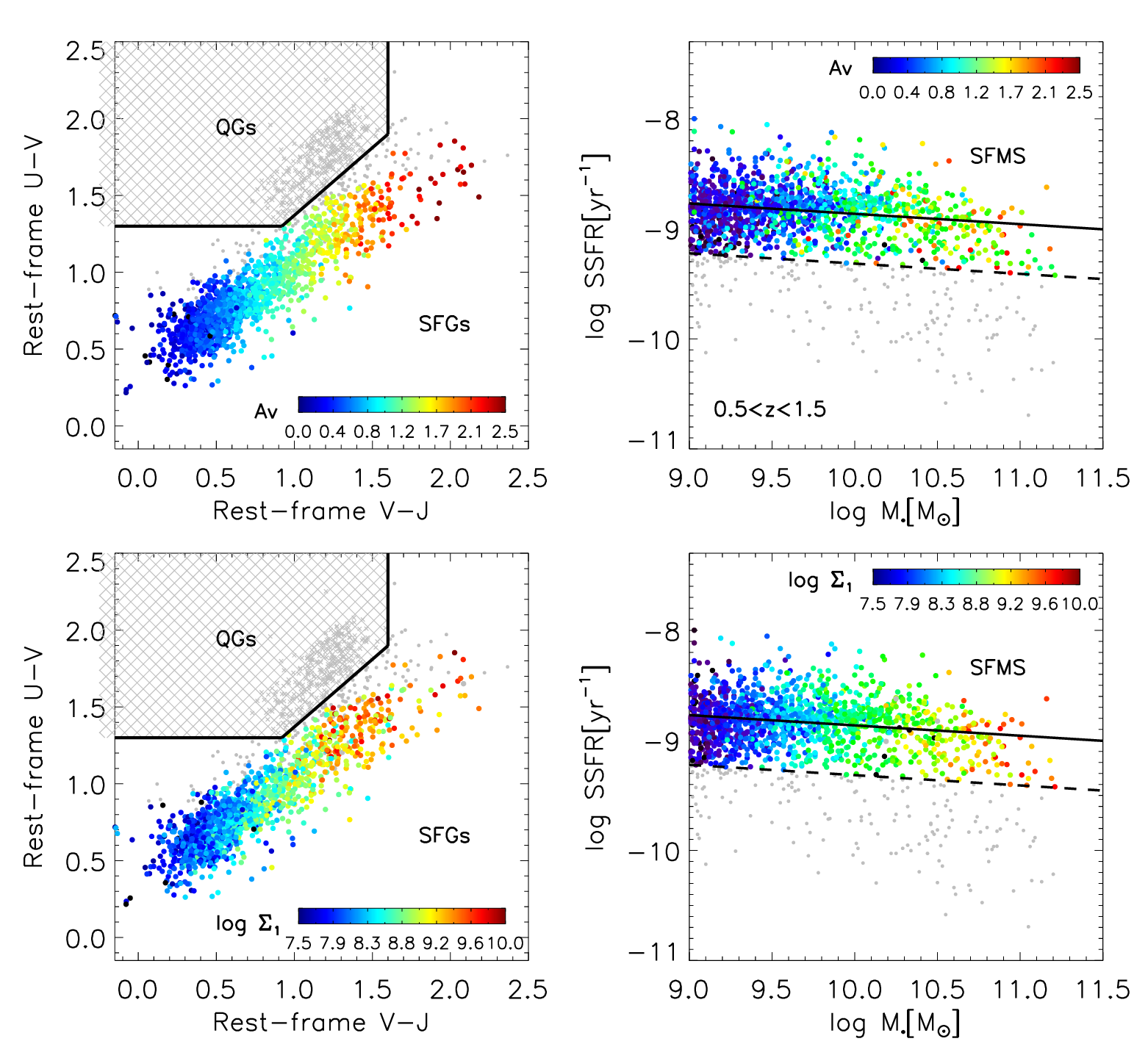}
\caption{
Rest-frame global $UVJ$ diagram (left) for our sample galaxies
after applying the selection cuts (see \S1) and SSFR vs. stellar mass (right)
for only $UVJ$-defined star-forming galaxies. Quiescent galaxies
are shown with gray hatching in the left panels.
The solid lines in the left panels indicate the selection criterion provided
by Williams et al. (2009).
The solid lines in the right panels show the best linear relation to the SFMS in our
adopted redshift range. The ``green-valley'' galaxies,
defined to have residuals $\rm \triangle log SSFR < -0.45~dex $
(below dashed lines) are shown with gray solid points.
Points for main-sequence SFGs are color-coded by
the global $\rm A_V$ (top panels) and $\rm {log \Sigma_{1}}$ (bottom panels),
respectively.
\label{sample_uvj}}
\end{figure*}

\begin{figure*}
\centering
\includegraphics[angle=0,width=1\textwidth]{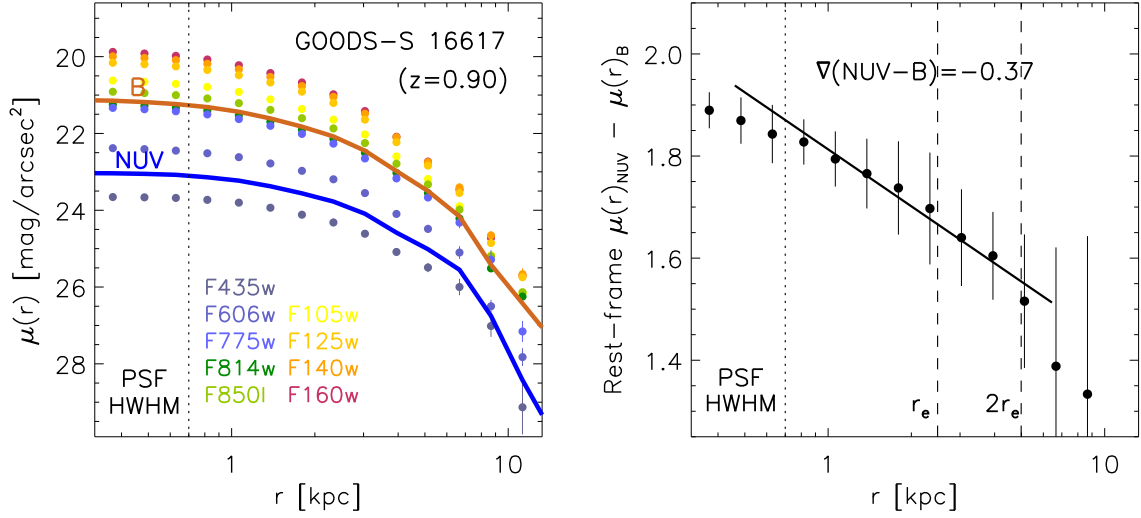}
\caption{
Example galaxy GOODS-S 16617 illustrating our measurement of rest-frame $NUV-B$ color gradient.
The observed {\it HST} multi-band surface brightness profiles and derived rest-frame $NUV$ and $B$ band
surface brightness profiles are shown in the left panel. Rest-frame $NUV-B$ color profile and
the best linear fit are shown in the right panel.
\label{relation}}
\end{figure*}

\begin{figure*}
\centering
\includegraphics[angle=0,width=0.95\textwidth]{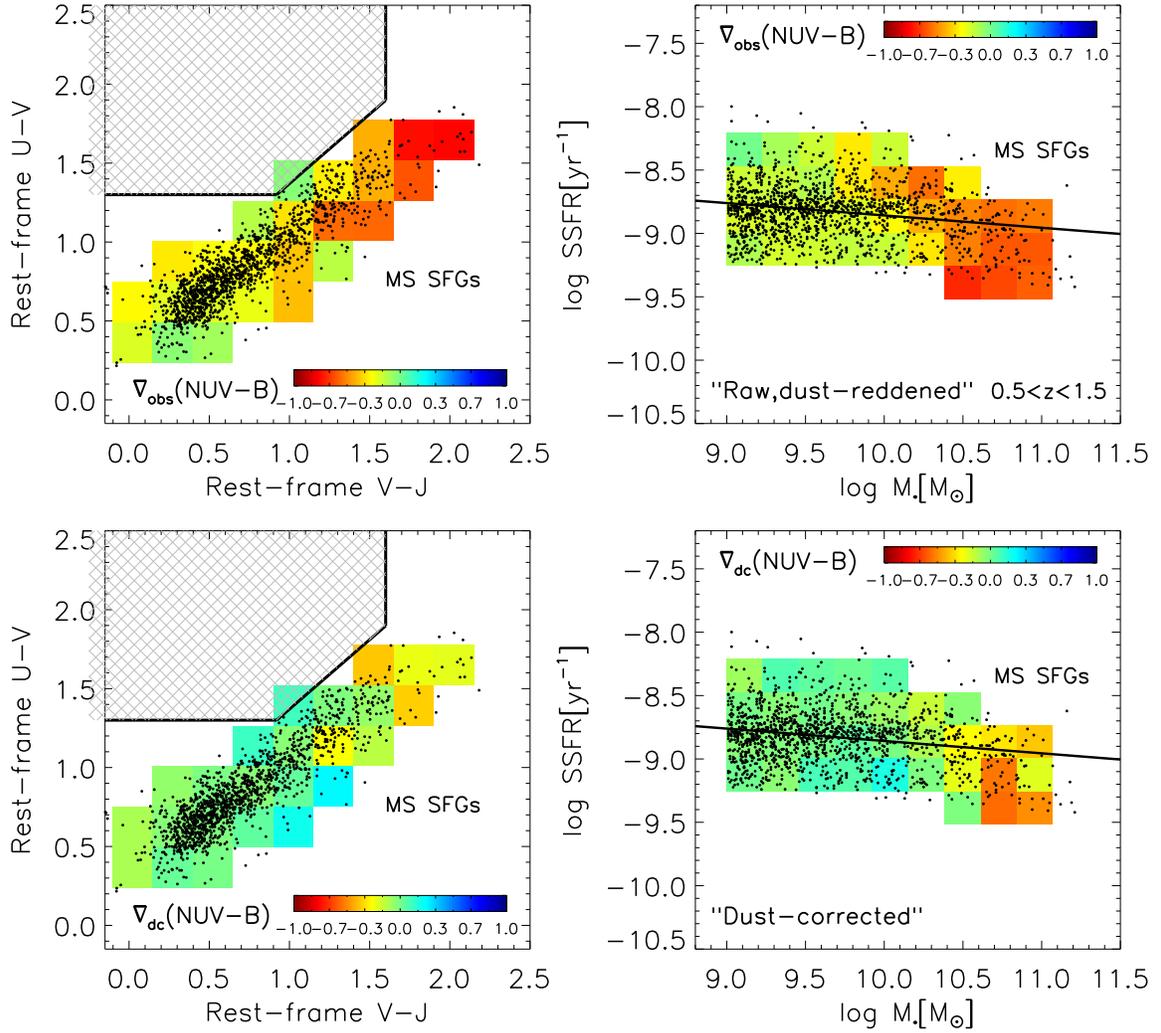}
\caption{
Rest-frame global $UVJ$ diagram (left) and SSFR vs. stellar mass (right) for
the main-sequence SFGs used in this study. Points are binned and color-coded by 
raw dust-reddened $NUV-B$ color gradients as observed (top panels) and
dust-corrected $NUV-B$ color gradients (bottom panels), respectively.
\label{relation}}
\end{figure*}

\begin{figure*}
\centering
\includegraphics[angle=0,width=0.96\textwidth]{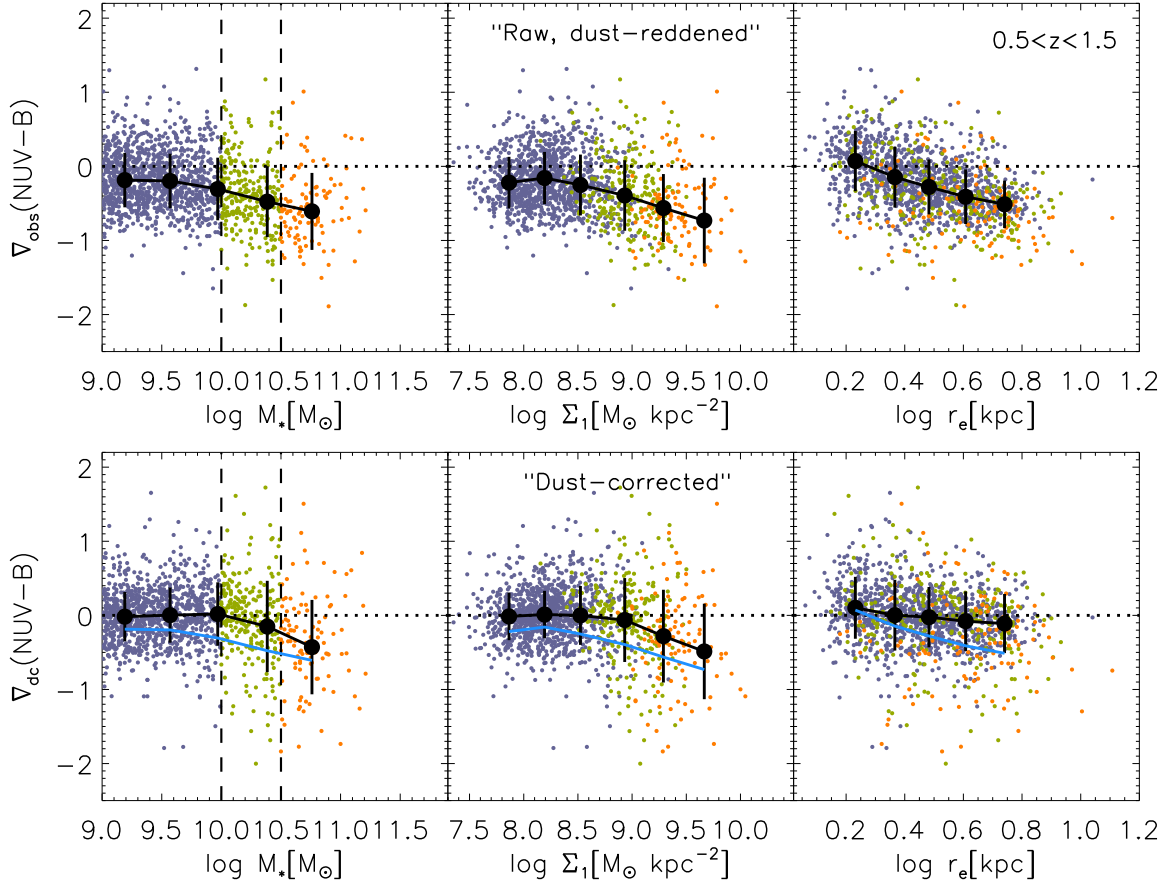}
\caption{
Raw dust-reddened (as observed, upper panels) and dust-corrected (bottom panels)
rest-frame $NUV-B$ color gradients in the main-sequence SFGs at $0.5<z<1.5$ as a function of
their stellar masses, $\rm{\Sigma_{1}}$ and rest-frame $V-J$ colors
from left to right, respectively.
The median values and corresponding standard deviations in classified bins
in each panel are shown with black symbols.
The trends for the correlations in the upper panels overlay the lower panels
with blue solid lines. The horizontal dotted lines indicate zero gradients.
The sample galaxies are divided into three mass subgroups ($M_{\ast} <10^{10}$$M_{\odot}$,
$10^{10}M_{\odot}<$$M_{\ast} <10^{10.5}$$M_{\odot}$
and $M_{\ast} >10^{10.5}$$M_{\odot}$), which are shown with different colors, respectively.
\label{relation}}
\end{figure*}

\end{document}